\documentclass[12pt]{article}
\usepackage{epsf}
\usepackage{amsfonts}
\usepackage{amsmath}

\newcommand{\beq}{\begin{equation}}
\newcommand{\beql}[1]{\begin{equation}\label{eq:#1}}
\newcommand{\eeq}{\end{equation}} 
\newcommand{\beqn}{\begin{eqnarray}}
\newcommand{\eeqn}{\end{eqnarray}}

\newcommand{\vev}[1]{\langle{#1}\rangle}

\newcommand{\field}[1]{\mathbb{#1}}
\newcommand{\BC}{{\field C}}

\newcommand{\BR}{{\field R}}
\newcommand{\ZZ}{{\field Z}}

\newcommand{\vQ}{{\cal Q}}\newcommand{\vbQ}{\overline{\vQ}}
\newcommand{\vL}{{\cal L}}
\newcommand{\bu}{\overline u}
\newcommand{\bd}{\overline d}
\newcommand{\be}{\overline e}
\newcommand{\bq}{\overline q}
\newcommand{\bH}{\overline H}
\newcommand{\bnu}{\overline \nu}
\newcommand{\de}[1]{\Delta_{#1}}

\newcommand{\myfig}[3]{\begin{figure}[ht]
\begin{center}
\leavevmode
\epsfxsize=#2cm
\epsfbox{#1}
\end{center}
\caption{#3}
\label{fig:#1}
\end{figure}}
\newcommand{\uiaddress}
{{\small\it Department of Physics, University of Illinois, Urbana, IL 61801}}
\newcommand{\email}[1]{\thanks{e-mail: \tt#1}}
\newcommand{\preprint}
{\begin{flushright}\begin{small}
ILL-(TH)-01-02\\ hep-ph/0105042\\ 
\end{small}\end{flushright}       
}

\begin{document}
\begin{titlepage}
	\title{\preprint\vspace{1.5cm}
               The Standard Model on a D-brane}
	\author{David Berenstein,\email{berenste@pobox.hep.uiuc.edu} \\
                Vishnu Jejjala,\email{vishnu@pobox.hep.uiuc.edu} \\ and \\     
                Robert G. Leigh\email{rgleigh@uiuc.edu}\\
        \uiaddress
\\
}
\maketitle

\begin{abstract}
We present a consistent string theory model which reproduces the Standard Model,
consisting of a $D3$-brane at a simple orbifold singularity. We study some simple
features of the phenomenology of the model. We find that the scale of stringy
physics must be in the multi-TeV range. There are natural hierarchies in the
fermion spectrum and there are several possible experimental signatures of the
model.\end{abstract}
\end{titlepage}

\section{Introduction}

A common thread in recent new proposals for physics beyond the Standard Model is
the realization of the gauge theory on a brane. In string theory terms, this is
presumably a D-brane. In this note, we will study a remarkably conservative
realization of the Standard Model in a fully consistent string background. The
local geometry is (a deformation of) the orbifold $\BR^4\times \BC^3/\Gamma$ with
the brane extended along the $\BR^4$. $\Gamma$ is a particular non-Abelian
discrete subgroup of $SU(3)$.

There are several interesting features present. First, there is a
natural hierarchy between the masses of leptons and quarks, because superpotential
lepton Yukawa couplings are forbidden by continuous gauge symmetries. We find,
however, that it is possible to achieve a realistic lepton mass spectrum through
K\"ahler potential terms after supersymmetry breaking. Assuming that these
nonrenormalizable terms are generated (at tree level) at the string scale, the
string scale must be in the multi-TeV range. We will not consider the global
geometry off the D-brane in detail here, but it should be noted that this
geometry must give rise to the TeV range string scale as well as supersymmetry breaking.
We will parametrize this breaking through effective spurion couplings in the
K\"ahler potential. Secondly, there are two additional gauged $U(1)$ symmetries
that are broken only at the weak scale. The phenomenology of these symmetries
deserves further study, but their presence does not seem to be in conflict with
experimental results.

\section{The Orbifold Model}

Consider a $D3$-brane at an isolated orbifold point in $\BC^3/\Gamma$, where
$\Gamma=\de{27}$, one of the non-Abelian discrete subgroups of $SU(3)$.  As such, the
resulting gauge theory has $N=1$ supersymmetry.  The group is one of the $\de{3n^2}$
series, defined by the short exact sequence
\begin{equation}
0 \rightarrow \ZZ_n \times \ZZ_n \rightarrow \de{3n^2} \rightarrow \ZZ_3 \rightarrow 0.
\end{equation} 
They are generated by three elements $e_1, e_2, e_3$ whose action on 
$\BC^3$ is given by
\begin{eqnarray}
e_1:&&(z_1,z_2,z_3) \rightarrow (\omega_n z_1,\omega_n^{-1} z_2,z_3), \nonumber \\
e_2:&&(z_1,z_2,z_3) \rightarrow (z_1,\omega_n z_2,\omega_n^{-1} z_3),\\
e_3:&&(z_1,z_2,z_3) \rightarrow (z_3,z_1,z_2), \nonumber
\end{eqnarray}
where $\omega_n$ is an $n$-th root of unity.  The quivers \cite{DM} of the
$\de{3n^2}$ groups are discussed in Refs. \cite{greene, muto1, muto2, HH,
bjl3}.  For the case $n=3$ that we are interested in, the quiver is as shown
in Figure
\ref{fig:Delta27two.eps}.  
\myfig{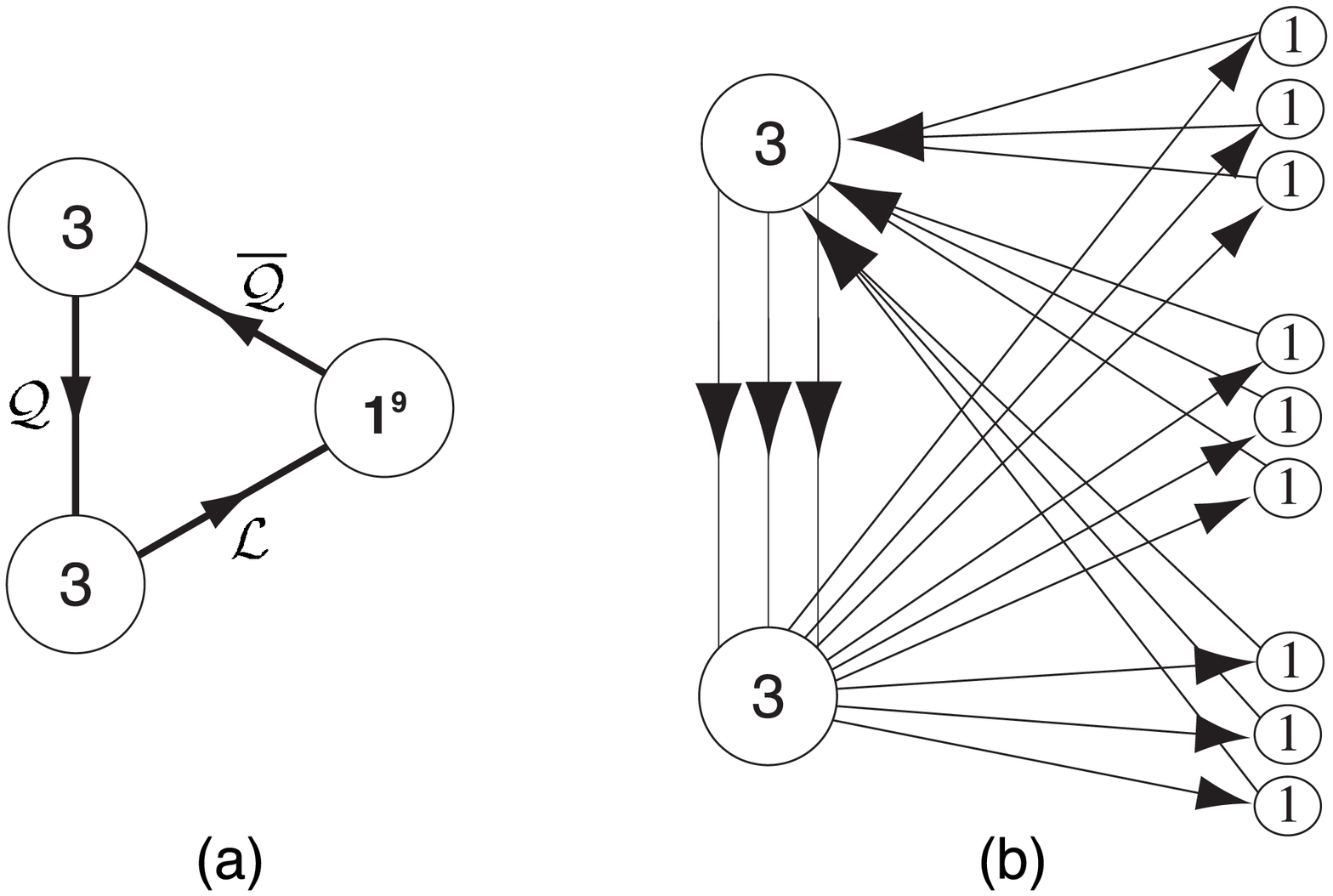}{7}{Two views of the quiver diagram of the $\de{27}$ singularity.} 
The gauge group is $\left(U(3)_+ \times U(3)_- \times U(1)^9\right)/U(1)$, and
the matter fields transform in the representations $\vQ_i=({\bf
3}_+,\overline{{\bf 3}}_-,0)$, $\vL_a=({\bf 1}_0,{\bf 3}_+,-_a)$, and
$\vbQ_a=(\overline{{\bf 3}}_-,{\bf 1}_0,+_a)$, where the index $a$ runs over the
nine $U(1)$'s and $i=1,\ldots,3$. The plus and minus subscripts denote the $U(1)$
charge under the decomposition $U(3) \sim SU(3) \times U(1)$.  Each of the fields
$\vL_a$ and $\vbQ_a$ are charged under only one of the nine $U(1)$'s. We will
identify the $SU(3)$ subgroup of $U(3)_+$ with the color group and $SU(2)_W$ is
embedded in the $U(3)_-$ group.  The orbifold theory comes with a renormalizable
superpotential generated at string tree-level of the form
\begin{equation}
W_0=\sum_{ia}\lambda_{ia} \vQ_i \vL_a \vbQ_a,
\end{equation}
where the $\lambda_{ia}$ are couplings of order one at the string scale. We will
study this superpotential in detail in what follows.

The orbifold has a number of moduli that we will exploit. There are two issues to
be addressed: first, at the orbifold point ({\it i.e.} all moduli vevs are zero),
the gauge couplings of the various group factors are related to one
another.\footnote{This situation was studied in Ref. \cite{aiqu} but the model
was rejected because $\sin^2 \theta_W$ is too small {\it at the orbifold point.}}
However, there are closed string moduli whose vevs shift the values of the
various gauge couplings, and so allowing for this, we may set the couplings as
needed. A second issue is that we wish to break some of the gauge symmetry: the
$U(3)_W$ should be broken to $SU(2)_W$, and at least some of the $U(1)$'s
removed. For the former, there are moduli corresponding to Fayet-Iliopoulos (FI)
terms $\int d^4\theta\, r_a V_a$ for the $U(1)$'s. The resulting $D$-term
equations are
\begin{equation} 
r_a-\vev{\vL_a^\dagger \vL_a}+\vev{\vbQ_a^\dagger \vbQ_a}=0,
\end{equation} 
and thus there are vacua where $\vev{\vL_a}\neq 0$. We will suppose that six of
the nine vevs are non-zero.\footnote{The other three fields will carry non-zero
hypercharge.} For clarity however, we will consider only three such vevs. This is
sufficient to display the structure of the non-Abelian symmetry breaking pattern,
and the inclusion of additional vevs will be accounted for later. Thus for now,
we will suppose that three of the FI terms are non-zero and positive,
$r_{1,2,3}>0$. Then the solutions $\vev{\vL_{1,2,3}}\neq 0$ may be chosen to
break $U(3)_W\times U(1)^3$ to $SU(2)_W\times U(1)_0$; $U(3)_c$ and the remaining
six $U(1)$'s are unbroken by these three vevs. Note that under this breaking pattern, we may
write (for
$i=1,2,3$)\footnote{We've replaced the index $a$ by a pair $i,j$.}
\begin{equation}
\begin{array}{ccclc}
\vQ_i & \to & Q_i & , &q_i\\
\vL_{1,i} & \to & L_i & , & g_i\\
\vL_{2,i} &\to & H_i &  , &\be_i\\
\vL_{3,i} &\to & \bH_i & , & \bnu_i\end{array}
\label{eq:fields}
\end{equation}
and we now make the identification
\begin{equation}
\vbQ_{1,i}\to\bq_i,\ \ \ \
\vbQ_{2,i}\to\bu_i,\ \ \ \ 
\vbQ_{3,i}\to\bd_i.
\end{equation}
The notation is that of the Standard Model, apart from the superfields $q, \bq,
g$. The fields $g_i$ are those that have vacuum expectation values. The
$q_j,\bq_j$ have a mass of order $\lambda_{ij}\vev{g}_i$ and may be integrated
out. Thus, we are left with the superfields of the three generation Standard
Model, including neutrino singlets, with six Higgs doublet superfields. With this
notation, the superpotential may be written
\begin{equation} 
W = \sum_{ij} \left\{ a_{ij} \left[ Q_i H_j + q_i \be_j \right] \bu_j + b_{ij} \left[ Q_i \bH_j + q_i \bnu_j \right] \bd_j + c_{ij} \left[ Q_i L_j + q_i {g_j} \right] \bq_j \right\},
\end{equation}
and in the broken phase becomes (unitary gauge)\footnote{To be precise, we have
$\tilde a_{ijk}=c_{ij}(\hat c^{-1}a)_{jk}$ and $\tilde b_{ijk}=c_{ij}(\hat
c^{-1}b)_{jk}$ where $\hat c_{ij}= c_{ij}\vev{g_j}$.}
\begin{equation}
W_{\rm eff} = 
\sum_{ij} \left\{ a_{ij} H_j Q_i \bu_j + b_{ij} \bH_j Q_i \bd_j - 
\tilde a_{ijk}Q_i\bu_kL_j\be_k-\tilde b_{ijk}Q_i\bd_kL_j\bnu_k\right\}.
\label{eq:Weff}
\end{equation}
We note that quark Yukawa couplings are present at tree level, but lepton Yukawas
are not. In addition, there are no $\mu$ terms, as all quadratic terms are
forbidden by gauge symmetries. 

Now, there are additional vevs that could be turned
on without further breaking the Standard Model gauge group. In the notation
presented here, these are the three sneutrinos $\bnu_j$. These vevs have several
virtues, primarily in that they break additional $U(1)$ symmetries, but also
that they are necessary, as we will see later, for realistic fermion masses. The
one thing that should be noted here is that if $\vev{\bnu}\neq0$, then there are
some field redefinitions that need to be done (e.g., in eq. (\ref{eq:Weff}) it can
be seen that $L$ mixes with $\bH$; there is also mixing with massive gauginos because of $SU(3)_-$ breaking).

\section{Anomalies and the Fate of the $U(1)$'s}

Let us discuss the unbroken gauge group in some detail. At the orbifold point
there are ten $U(1)$'s.  We find that there are the following non-zero gauge
anomalies
\begin{eqnarray}
U(3)_\pm U(3)_\pm U(1)_\mp: && \mp 3\sqrt{3}, \nonumber \\
U(3)_\pm U(3)_\pm U(1)_a: && \pm \sqrt{3}, \label{eq:anomalies} \\
U(1)_a U(1)_a U(1)_\pm: && \mp 3\sqrt{3}. \nonumber
\end{eqnarray}
There are no gravitational or mixed anomalies. The generalized Green-Schwarz
mechanism (involving the NSNS B-field as well as twisted
moduli)\cite{GS,GSW2,DM,BLPSSW} will serve to cancel these anomalies and
consequently will break some of the $U(1)$'s. The generators of the broken
$U(1)$'s are $3Y_\pm \mp \sum_a Y_a$. Since the $U(1)$ generated by
$Y_++Y_-+\sum_aY_a$ decouples, we can conclude that all surviving generators are
orthogonal to $Y_\pm$ as well as $\sum_a Y_a$. Note also that since baryon number
may be identified with $U(1)_+$, there are no perturbative baryon number
violating processes such as proton decay, as the global symmetry survives the
Green-Schwarz mechanism; see also \cite{Anton} where the same mechanism was
realized in a different model.

Now, if we consider the $g$ and $\bnu$ vevs, there is an unbroken $U(1)$
generator which we call $Q_0$; this is a linear combination of $Q_{8-}$ (the
diagonal generator of $SU(3)_-$) and six of the nine $U(1)$'s:
\begin{equation}
Q_0=\sqrt{6}Q_{8-}-2\sum_j
(Q_{1,j}+Q_{3,j}).
\end{equation}
This can be mixed with a linear combination of the $Q_{2,j}$'s as long as we
take a non-anomalous combination. Thus, we identify the hypercharge
generator
\begin{equation}
Y =-\left[Q_0 + 4 \sum_{i=1}^3 \left(  Q_{2,i} \right) \right]/6.
\label{eq:hypercharge}
\end{equation}
The $U(1)$ charges in the low energy theory are tabulated below.
\begin{equation}
\begin{array}{|c||r|r|r|r|r|}
\hline
& U(1)_0 & U(1)_1 & U(1)_2 & U(1)_3 \\ \hline \hline
Q & -1 & 0 & 0 & 0  \\ \hline
\bu & 0 & 0 & 1 & 0  \\ \hline
\bd & -2 & 0 & 0 & 1  \\ \hline
L & 3 & -1 & 0 & 0  \\ \hline
\be & -2 & 0 & -1 & 0  \\ \hline
\bnu & 0 & 0 & 0 & -1  \\ \hline
H & 1 & 0 & -1 & 0  \\ \hline
\bH & 3 & 0 & 0 & -1  \\ \hline
g & 0 & -1 & 0 & 0 \\ \hline
q & 2 & 0 & 0 & 0 \\
\hline\bq & -2 & 1 & 0 & 0 \\
\hline
\end{array}
\label{eq:charges}
\end{equation}
There are two other unbroken non-anomalous $U(1)$'s present other than
hypercharge. We can take these to be $Q_{2,1}-Q_{2,3}$ and $Q_{2,2}-Q_{2,3}$.
Under these two symmetries, only $\bu_j, \be_j$ and $H_j$ are charged, as
can be seen by looking at the table. We will comment later on the possibly
interesting phenomenology associated with these extra $U(1)$ symmetries, which are
broken at the weak scale.

\section{Mass Spectrum, Couplings  and Supersymmetry \\ Breaking}

{}From the superpotential, eq. (\ref{eq:Weff}), we see that the quark sector
has standard Yukawa couplings giving rise to supersymmetric mass terms
\begin{equation}
(m_u)_{ij}=a_{ij}\vev{H_j},\ \ \ \ \ \ (m_d)_{ij}=b_{ij}\vev{\bH_j}.
\end{equation}
There are no Yukawa couplings present for the leptons, a consequence of
the gauged $U(1)$ symmetries of the orbifold. We regard this as a strong
feature of the model: the lepton masses are hierarchically suppressed
compared to quark masses. We need to demonstrate of course that lepton
masses can be generated. Because of the symmetries, the lepton masses
must be generated through K\"ahler potential terms, as follows. There will
be terms of the general form
\begin{equation}
K\supset \frac{1}{M^2} \alpha_{ab} (\vL_a^\dagger\vL_b)(\vL^\dagger_b\vL_a)+
\frac{1}{M^2} \alpha'_{ab} (\vL_a^\dagger\vL_a)(\vL^\dagger_b\vL_b).
\end{equation}
In particular, we will find terms
\begin{equation}
\alpha_{ij} g_i^\dagger\be_j H_j^\dagger L_i +
\beta_{ij} g_i^\dagger\bnu_j
\bH^\dagger_j L_i
\end{equation}
which give rise to charged lepton fermion masses of the
form\begin{equation}(m_L)_{ij}\sim \alpha_{ij}
\frac{F_{g_i}^*}{M^2}\vev{H^*_j}\end{equation} and neutrino Dirac masses
\begin{equation}
(m_D)_{ij}\sim \beta_{ij}
\frac{F_{g_i}^*}{M^2}\vev{\bH^*_j}.
\end{equation}
From these equations it is clear that the generation of lepton Yukawas are
intimately tied with the supersymmetry breaking scale and each of them is
hierarchically suppressed.

Supersymmetry may be broken in a variety of ways in brane models. We will
take an agnostic approach, and simply write the effects of supersymmetry
breaking in terms of spurion couplings. For example, we will take the
K\"ahler potential to contain terms
\begin{equation}
K=\ldots+\frac{1}{M} S\sum 
\phi_i^\dagger\phi_i+\frac{1}{M^2} \Psi^\dagger\Psi \phi_i^\dagger\phi_i
+\ldots,\label{eq:spurions}
\end{equation}
where the $\phi_i$ are any of the open string modes. The spurions $S$
and $\Psi$ may very well be closed string modes,\footnote{For example,
$S$ can be identified, more or less, with the dilaton.} which we would
expect to couple universally to the open string modes. We will make the
assumption that $\vev\Psi\sim \theta^2 F\sim\vev{S}$, so that
$m_{susy}\sim F/M$. The scale $M$ is some high energy scale, which we
identify with the string scale. Note that with the K\"ahler potential
given, we find that open string fields may have $F$-terms of the
form
\begin{equation}F_i\sim \frac{F\vev{\phi_i}}{ M}\sim
m_{susy}\vev{\phi_i}.\label{eq:visvev}
\end{equation} 
Under these
conditions the lepton masses are suppressed by a factor
$m_{susy}\vev{g}/M^2$ according to (\ref{eq:visvev}). For this reason
$M$ cannot be too far above the supersymmetry breaking scale $m_{susy}$.
A possible scenario is where $M\sim$ 10 TeV, $\vev{g}\sim\vev{\bnu}\sim$
1 TeV, $m_{susy}\sim$ 3 TeV. Of course, in order to obtain a realistic
spectrum, the neutrino masses must be suppressed compared to the charged
leptons.  By the same mechanism, there are also neutrino Majorana masses
of the form
\begin{equation} 
\gamma_{ij} \frac{F_{\bnu_i}^*\vev{\bnu^*_j}}{M^2}
\end{equation}
Comparing to the Dirac masses, we see that these are larger, and thus a
seesaw can occur. Roughly, we have
\begin{equation}\frac{m_\nu}{m_L}\sim 
\frac{\beta^2}{\alpha\gamma}\frac{\vev{H}\vev{g}}{\vev{\bnu}^2}\left(\frac{\vev{\bH}}
{\vev{H}}\right)^2.\end{equation}
The simplest way that the mass hierarchy between up and down type quarks
may be obtained is to take $\vev{\bH}<\vev{H}$ (large $\tan\beta$) and
we see that the light neutrino masses are suppressed by two powers of
$\tan\beta$. This is probably not sufficient, but note that there are
several other mechanisms available here to suppress neutrino masses.
First, there may be small differences in coupling constants, and
possibly a difference between $\vev{g}$ and $\vev{\bnu}$. As well, since
there are six Higgs doublets, there may be a hierarchy of vevs which
could further suppress low generation neutrino masses. We conclude that
there is ample parameter space to obtain a realistic lepton mass and
mixing spectrum.

All superpartners receive diagonal supersymmetry breaking masses of
order $m_{susy}$, since the scalar masses coming from
(\ref{eq:spurions}) are universal. There are also off-diagonal masses
which would come about through K\"ahler potential terms of the form
\begin{equation} 
K\supset \frac{1}{M^2}\alpha_{ij,a}(\vQ^\dagger_i\vQ_j)(\vL^\dagger_a\vL_a)+
\beta_{ab}(\vbQ^\dagger_a\vbQ_a)(\vL^\dagger_b \vL_b) +
\gamma_{ab}(\vbQ^\dagger_a\vbQ_b)(\vL^\dagger_a \vL_b).\label{eq:fcnc}
\end{equation}
Thus, there are off-diagonal scalar masses, for example
\begin{equation}
(m_Q^2)_{ij}\sim \frac{\alpha_{ij,a}|F_a|^2}{M^2}
\sim \frac{\alpha_{ij,a}|\vev{g_a}|^2}{M^2}m_{susy}^2.
\end{equation}
As a result, there is some suppression here of flavor changing neutral
currents. We have not done a complete analysis, but it appears that they
may be at or below experimental limits.

Notice also that each of the right-handed quarks couples to a different
Higgs multiplet. Hence one can keep all couplings of the same order of
magnitude and produce  the mass hierarchy between generations by having 
a hierarchy of vevs. Also, the CKM matrix may turn out to be nearly
diagonal because of approximate discrete symmetries that can appear near
the orbifold point.  This same discrete symmetry might account for the
hierarchy between the generations.

\section{Discussion}

We have presented here a consistent string model giving rise in the low
energy limit to a gauge theory which closely resembles the Standard
Model. The phenomenology of the model is rich. Lepton masses are
hierarchically suppressed compared to quark masses, and it appears, at
least at the level of analysis done here, that a realistic spectrum is
possible. An important aspect of the phenomenology is that we must
arrange that the string scale is low, thus the Planck mass in four
dimensions can be attributed to large extra dimensions \cite{HDD,AAHDD} or a
Randall-Sundrum type model \cite{RS}. A possible scenario is where
$M\sim$ 10 TeV, $\vev{g}\sim\vev{\bnu}\sim$ 1 TeV, $m_{susy}\sim$ 3 TeV.
The model possesses six Higgs fields, a pair for each generation, and
thus there is some flexibility in the fermion spectrum. We have not dealt with
the details of the Higgs spectrum; several of these may be heavy because of supersymmetry breaking effects, there is mixing with sleptons and $SU(3)_-$ breaking removes one
linear combination from the low energy spectrum.
The spectrum of
neutrino masses is also an interesting feature---it would be interesting
to explore this further in light of present experiments. There are
several additional aspects of the phenomenology that deserve more
careful consideration. These include flavor changing neutral currents,
and the presence of relatively light extra gauge bosons. These
symmetries are broken at the weak scale by Higgs vevs, but it should be
noted that these gauge bosons do not have weak-style couplings, and
hence the experimental constraints are not immediately apparent.

\bigskip\bigskip

\noindent {\bf Acknowledgments:} Discussions with S. Willenbrock are
gratefully acknowledged.  Supported in part by U.S. Department of Energy,
grant DE-FG02-91ER40677.  VJ is supported by a GAANN fellowship from the
U.S. Department of Education, grant 1533616.


\providecommand{\href}[2]{#2}\begingroup\raggedright\endgroup


\end{document}